\documentclass[twocolumn,showpacs,preprintnumbers,amsmath,amssymb]{revtex4}

\usepackage{graphicx}
\usepackage{dcolumn}
\usepackage{bm}

\begin{document}

\preprint{APS/123-QED}

\title{Extreme resistance of super-hydrophobic surfaces to impalement: reversible electrowetting related to the impacting/bouncing drop test} 

\author{P. Brunet$^1$, F. Lapierre$^2$, V. Thomy$^2$, Y. Coffinier$^3$ and R. Boukherroub$^3$}
\affiliation{$^1$Laboratoire de M\'ecanique de Lille, UMR CNRS 8107, Bd. Paul Langevin, 59655 Villeneuve d'Ascq, France\\
$^2$Institut d'Electronique de Micro\'electronique et de Nanotechnologies, UMR CNRS 8520, Cit\'e Scientifique, Avenue Poincar\'e, BP. 60069, 59652 Villeneuve d'Ascq, France\\
$^3$Institut de Recherche Interdisciplinaire (IRI), USR 3078 and IEMN, UMR CNRS-8520,
Cit\'e Scientifique, Avenue Poincar\'e, BP. 60069, 59652 Villeneuve d'Ascq, France
}

\date{\today}

\begin{abstract}

The paper reports on the comparison of the wetting properties of super-hydrophobic silicon nanowires (NWs), using drop impact impalement and electrowetting (EW) experiments. A correlation between the resistance to impalement on both EW and drop impact is shown. From the results, it is evident that when increasing the length and density of NWs: (i) the thresholds for drop impact and EW irreversibility increase (ii) the contact-angle hysteresis after impalement decreases. This suggests that the structure of the NWs network could allow for partial impalement, hence preserving the reversibility, and that EW acts the same way as an external pressure. The most robust of our surfaces show a threshold to impalement higher than 35 kPa, while most of the super-hydrophobic surfaces tested so far have impalement threshold smaller than 10 kPa.

\end{abstract}

\pacs{Valid PACS appear here}

\maketitle

\textit{Introduction} - The superhydrophobic (SH) character of a surface generally arises from an interplay between a rough structuration of the surface and an ad-hoc chemical treatment. The recent upsurge of interest on the subject \cite{Lau03,Patankar04} is supported by the development of more and more sophisticated techniques for micro- and nano-texturation, which make it possible to mimic water-repellent biosurfaces such as lotus leaves. One of the most striking behavior of such surfaces is that water drops slide or roll on them with almost no friction. This asset can be used to design lab-on-chip devices, in order to easily handle micro-droplets. In this sense, Electrowetting (EW) is one of the most promising techniques to carry out elementary operations on droplets, like displacement, splitting or merging \cite{Mugele_Baret05,MugeleReview,Verplank2}, but these operations can often be limited by an irreversible behavior on rough textured surfaces \cite{ew1,ew2,ew3}. In a recent report, we have shown that EW on SH surfaces could be realized in a reversible fashion \cite{Verplank_etal07}.

\begin{figure}
\begin{center}
\includegraphics[scale=0.35]{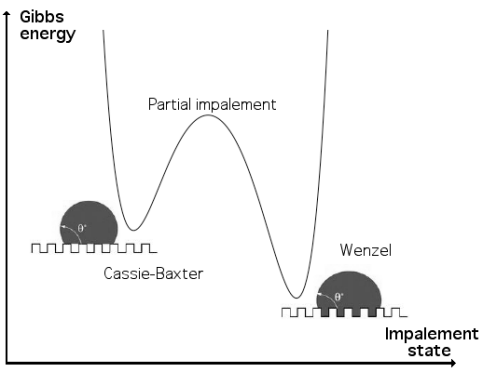}
\caption{A schematic view of the Gibbs free energy of the Cassie-Baxter and Wenzel states along the impalement configuration.}
\label{fig:energy}
\end{center}
\end{figure}

On SH surfaces, a drop reaches apparent contact angles (CA) higher than 150$^{\circ}$, whereas the corresponding non-textured - smooth - surfaces generally have (Young's) CA of about $\theta_0$ = 100$^{\circ}$. More than 50 years ago, Cassie and Baxter \cite{Cassie_Baxter} on one side, and Wenzel \cite{Wenzel} on the other side determined the apparent CA of a textured surface. The Cassie-Baxter's approach assumes that air pockets are trapped between the drop and the substrate, and that the drop partially sits on the emerging texture. Hence, the drop mainly sees air below, and adopts an almost spherical shape. In this case, the apparent CA is determined as follows:

\begin{equation}
\cos \theta = \phi_s (\cos \theta_0 +1) - 1 
\label{eq:cassie_baxter}
\end{equation}

\noindent where $\phi_s$ is the surface fraction of the liquid-solid interface. The Wenzel's approach assumes that the liquid fills all the pores. Under this assumption, the apparent CA equals:

\begin{equation}
\cos \theta = r \cos \theta_0
\label{eq:wenzel}
\end{equation}

\noindent where $r$ is the roughness of the surface, i.e. the ratio between the total surface and the projected one ($r \ge$ 1). From energy considerations \cite{QuereReview05}, it has been shown that the Cassie-Baxter state is often metastable. In practice, it means that the drop will remain in a Cassie-Baxter state only if it is subjected to soft external perturbations (see a schematic view in terms of the Gibbs free energy in Fig. \ref{fig:energy}). Sometimes, the weight of the drop itself may be sufficient to provoke the impalement.

Due to their huge values of contact angle hysteresis (CAH) \cite{Lafuma_Quere03,QuereReview05}, Wenzel states are not suitable for applications of low-friction drop displacement or self-cleaning : it is thus important to predict the transition between Cassie-Baxter and Wenzel states. Intuitively, the pressure threshold should be related to the pressure needed to fill the pores of the texturation: for a single-sized structure of pillars of size $d$, this pressure is in the order $P \simeq \frac{2 \sigma \cos \theta_0}{d}$, according to the Washburn equation, where $\sigma$ is the liquid surface tension. In practice, it has been shown \cite{Lafuma_Quere03} that this pressure is overestimated: the impalement occurs at much lower pressure $P_c$. To increase $P_c$, hence the resistance of the Cassie-Baxter state, a multiple-scale roughness has been proposed, inspired from the structure of lotus leaves \cite{Patankar04,Gao_McCarthy06,Bhushan_Nosonovsky06,Nosonovsky_Bhushan07}.

\begin{figure}
\begin{center}
(a)\includegraphics[scale=0.80]{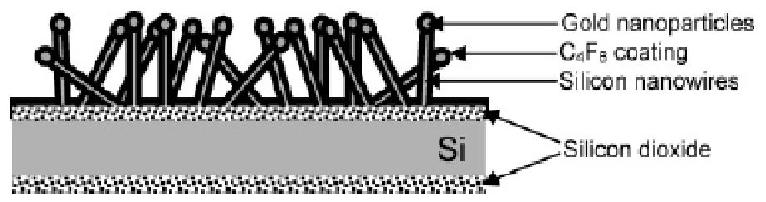}
(b)\includegraphics[scale=0.50]{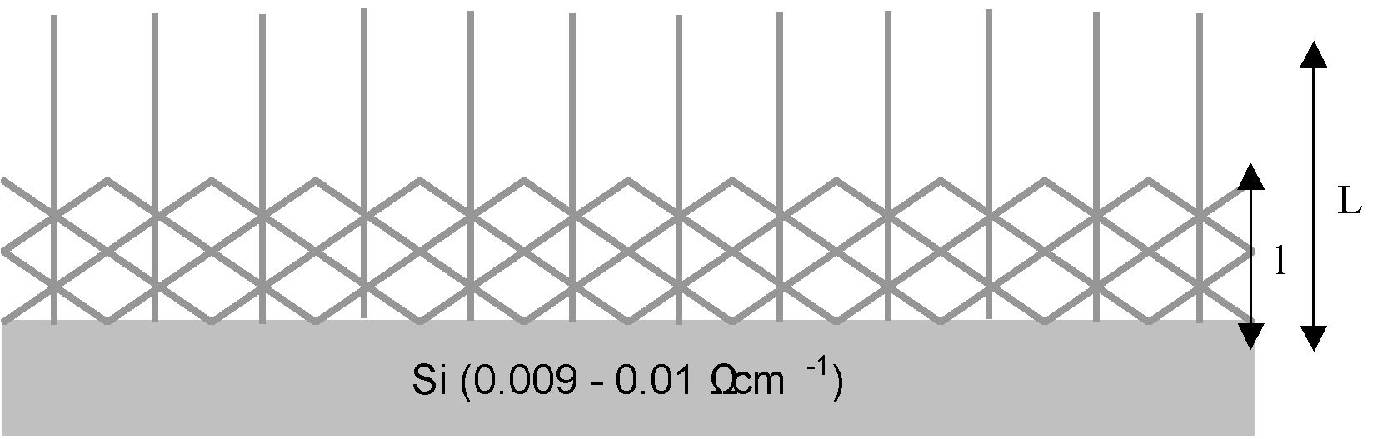}
\caption{(a) Schematic illustration of the nanotexturation. (b) Scheme of the resulting double texturation: an upper layer of straight NWs and a lower layer of crossed NWs.}
\label{fig:coating_scheme}
\end{center}
\end{figure}

\begin{figure}
\begin{center}
(a)\includegraphics[scale=0.15]{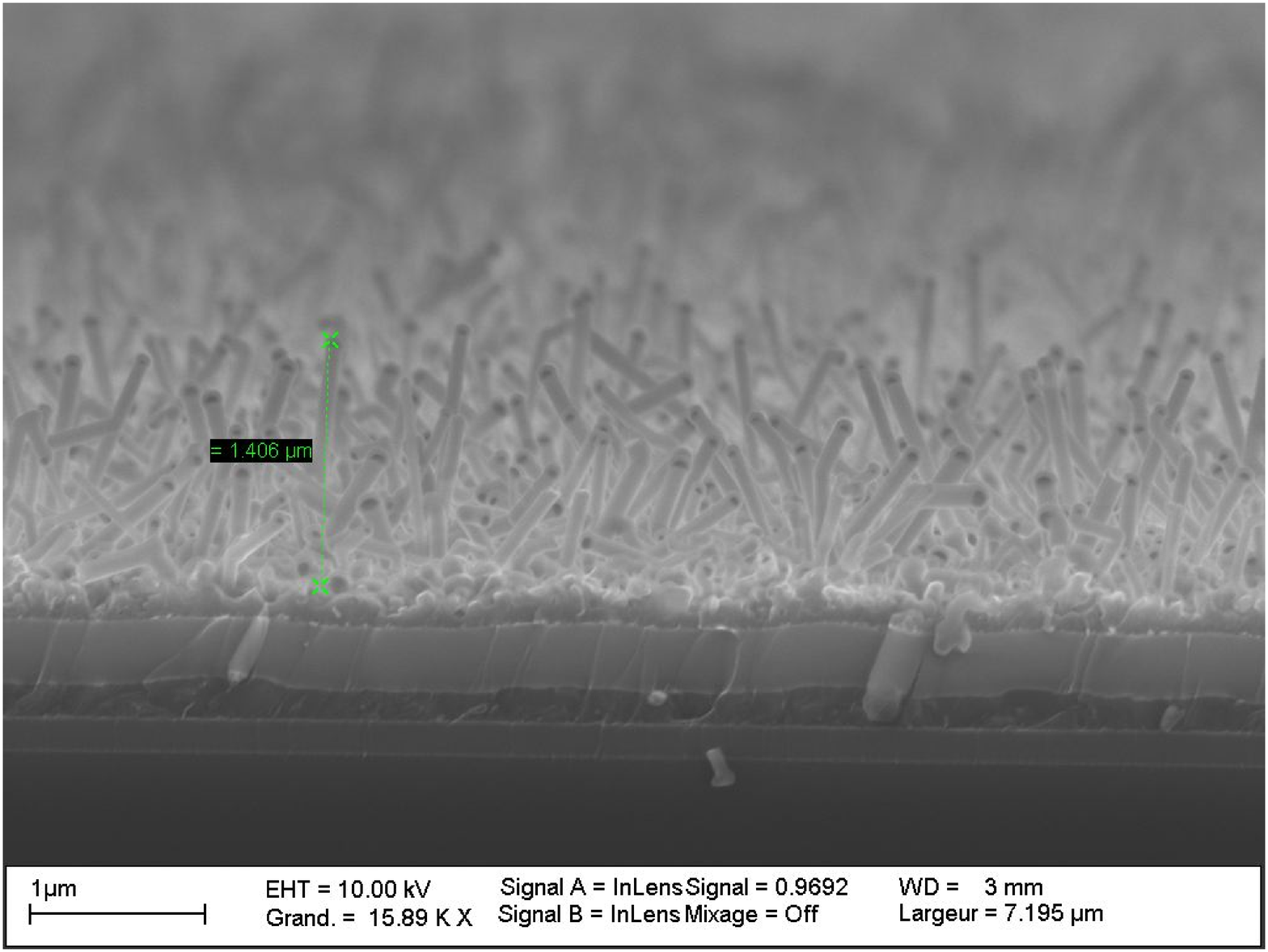}
(b)\includegraphics[scale=0.15]{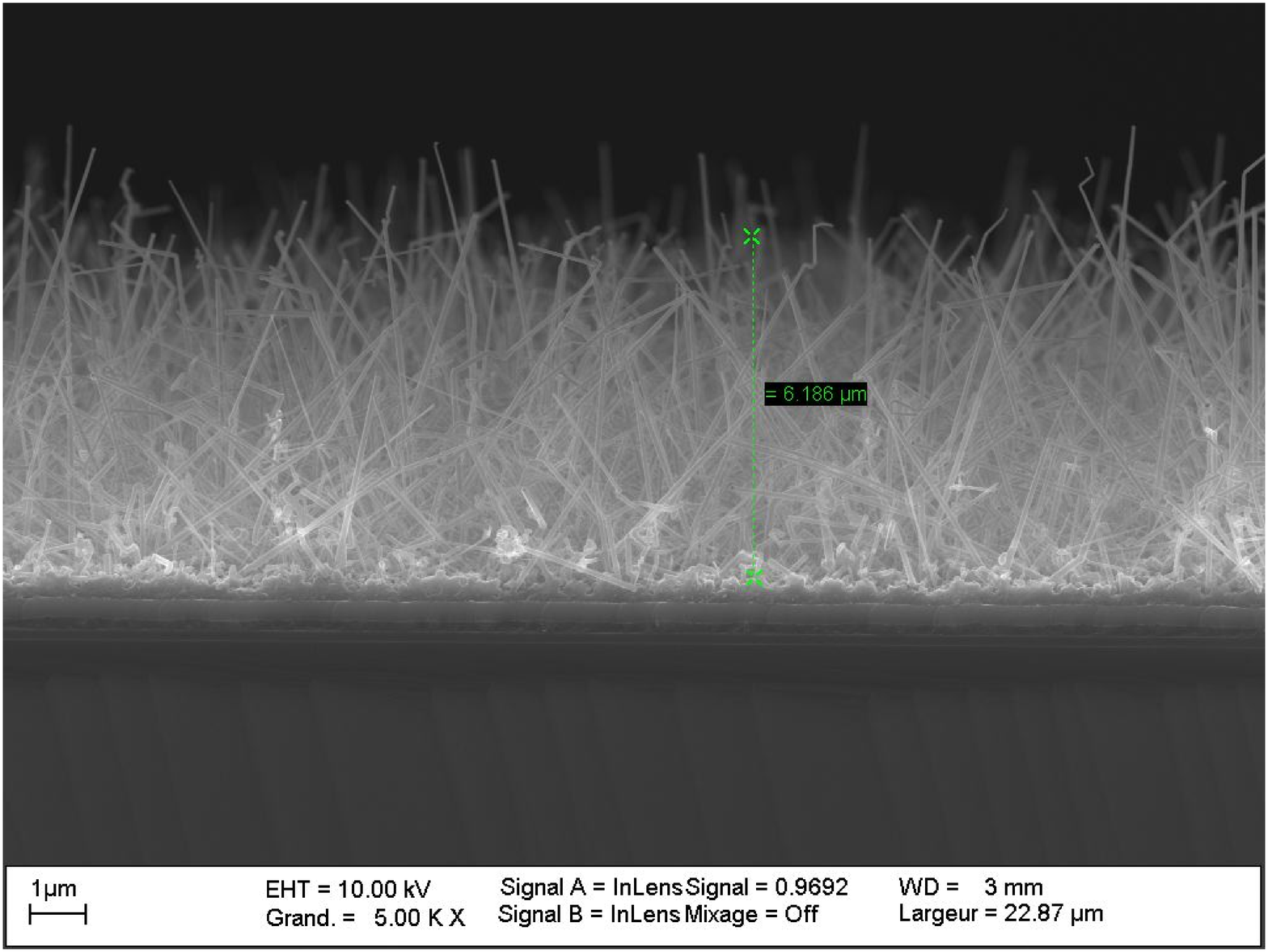}
(c)\includegraphics[scale=0.15]{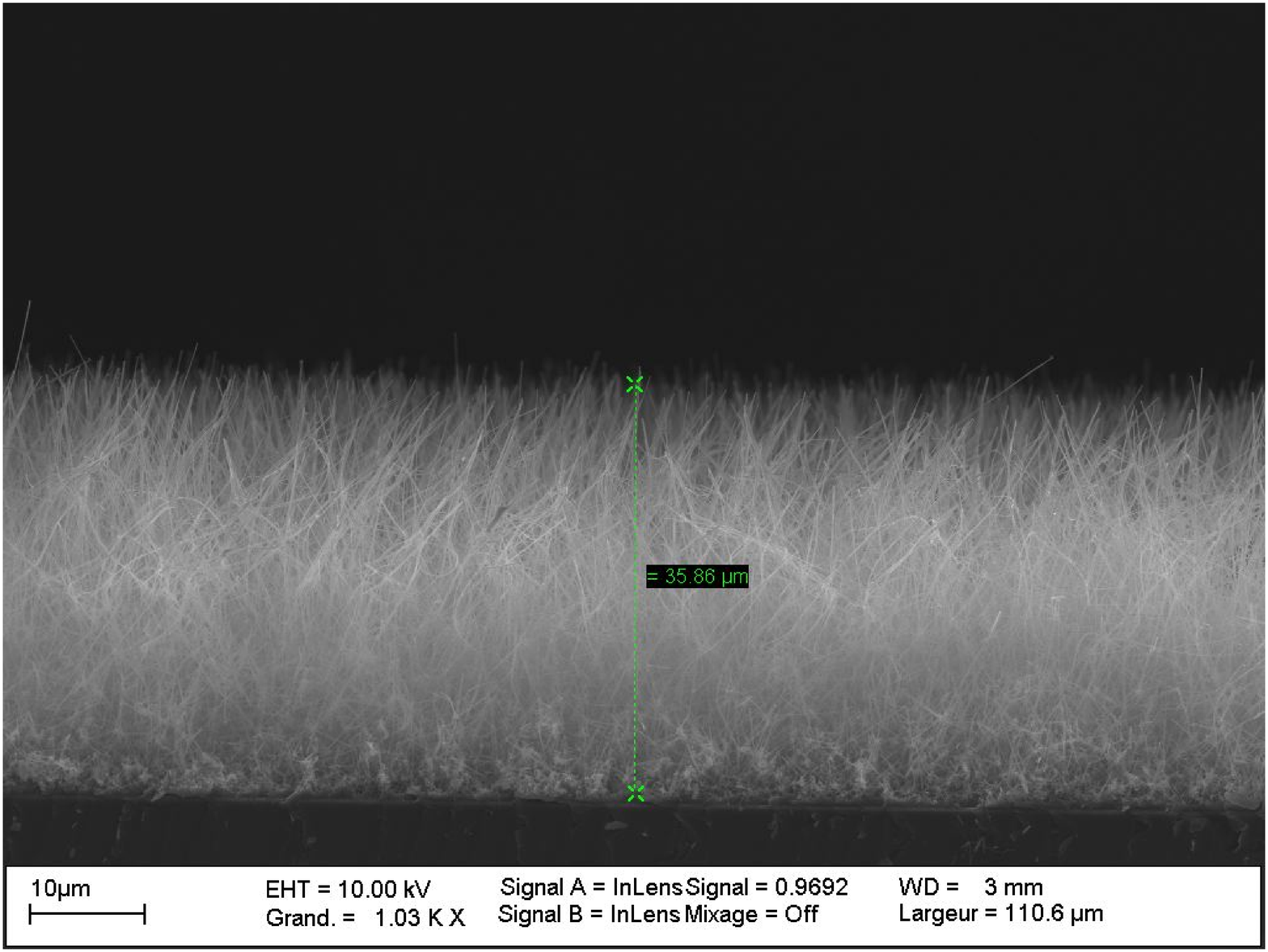}
\caption{(a-c) SEM images of the nano-wires carpet coated with $C_4F_8$, corresponding to P1, P2 and P3, respectively.}
\label{fig:coating}
\end{center}
\end{figure}

The work investigated in this paper uses a slightly different strategy for the design of robust silicon NWs surfaces. Based on our previous results on EW \cite{Verplank_etal07} using the VLS growth technique catalysed by gold nano-particles (see Fig. \ref{fig:coating_scheme}-a), we show that both the density and length of NWs are crucial to obtain a high resistance to impalement. SEM pictures of some of our surfaces are depicted in Fig. \ref{fig:coating}. The resistance to impalement is tested using drop impact experiments, a method which enables to work under high pressure, compared to other ways of measurements - like the squeezed drop between two parallel plates \cite{Lafuma_Quere03}. Similar experiments as well as tentative explanations have been recently undertaken by two groups \cite{Bartolo06,Quere_Reyssat06}. Their study was carried out on micro-textured surfaces, and a scaling-law was proposed for the threshold pressure: $P_c \sim \frac{\sigma h}{l^2} $, here $h$ is the height of the pillars and $l$ the distance between them. However, the reasons for this scaling and the value of the pre-factor are still under debate. A recent paper \cite{Moulinet_Bartolo} showed the existence of partial impalement of liquid inside arrays of micro-pillars, which could be a key for a fundamental understanding of the Cassie-Baxter to Wenzel transition. In short, they evidenced intermediate states between Cassie and Wenzel, where the liquid either protrudes the texture at intermediate height. A more recent paper by Kusumaatmaja et al. \cite{Kusumaatmaja08} analyses theoretically these intermediate states.

Here, we do not pretend to solve these points, but rather to show that the specific design of our surfaces, ensures a large threshold for the transition from Cassie-Baxter to Wenzel state, compared to the micron-sized textures. 
Hence, the purpose of the present paper is to show that the reversibility of EW is related to robustness to impalement, by comparing EW experiments to drop impact experiments on surfaces designed from a carpet of entangled or straight nano-wires (NWs). 

\textit{Preparation of superhydrophobic substrates} - 
Single-side polished silicon (100) oriented n-type wafers (Siltronix, France) (phosphorus-doped, 5-10 Ohm.cm$^{-1}$ resistivity) were used as substrate. The silicon substrate was first degreased in acetone and isopropyl alcohol, rinsed with Milli-Q water, and then cleaned in a piranha solution (3:1 concentrated H$_2$SO$_4$ /  30$\%$ H$_2$O$_2$) for 15 min at 80$^{\circ}$C followed by copious rinsing with Milli-Q water. The surface was further dried under a stream of nitrogen. 
A dielectric layer (SiO$_2$) of 300 nm thickness was grown on both side of the silicon substrate by thermal oxidation. 
Silicon NWs used in this work were prepared using the vapor-liquid-solid (VLS) mechanism as described in \cite{Verplank_etal07}. The fundamental process is based on metal-catalyst-directed chemical vapor deposition of silicon. First, a thin film of gold was evaporated on one side of the clean Si/SiO$_2$ substrate, serving as catalysts for the NWs growth. After the growing step, a surface composed of silicon NWs electrically insulated with 300 silicon dioxide layer was obtained. The SH was achieved through coating of the NWs with C$_4$F$_8$ by plasma deposition. Both NWs and gold nanoparticles are covered with the fluoropolymer. 

The three different processes presented in this paper are denoted P1, P2 and P3. P1 shows a low density of NWs with a length of 1 $\mu$m (Fig.\ref{fig:coating}-a).  P2 displays an average density with a disorganized layer of NWs (average length 7 $\mu$m, Fig.\ref{fig:coating}-b). Finally, P3 exhibits a two-layered structure: a dense layer of NWs with a height of 20 $\mu$m and a second upper layer with a low density and a height of 15 $\mu$m  (Fig.\ref{fig:coating}-c). Schematically, the resulting nano-textured surface is shown in Fig. \ref{fig:coating_scheme} for substrates of the same type as P3: a lower layer of crossed NWs (height $l$) and an upper layer of straight NWs (height $L-l$).

By changing the growth parameters (pressure, time), we obtained surfaces with different morphologies (Table I). Under peculiar conditions, the height of one of the layers (Fig. \ref{fig:coating_scheme}-b) can be negligible. Among the ten different superhydrophobic surfaces realized and characterized, only the three which displayed the most significant geometrical properties (height of the nanowires, density and double -layer) are discussed here. From all the ten surfaces tested, qualitative trends showed that the threshold for impalement was indeed increasing with both the length of the upper straight nanowires layer and the density of the lower cross-linked NWs. However, the three selected surfaces lead to three very different impalement behaviors.

\begin{table}
\begin{center}
\begin{tabular}{|c|ccc|}
\hline
Process 	 & Time (min) 	& Pressure (T)  & Height ($l$ / $L$) ($\mu$m)\\
\hline
P1		 & 	10 		& 	0,4 		& 1 / 1 \\
P2		 & 	60 		& 	0,1 		& 7 / 7 \\
P3		 &	60		& 	0,4		& 20 / 35 \\
\hline
\end{tabular}
\caption{VLS nano-wires Grown parameter on Si wafer}
\end{center}
\label{grown_parameter}
\end{table}

\textit{Drop impact experiments} - We used a dripping faucet that release a drop of liquid (water/glycerin mixture) from a sub-millimetric nozzle. The drop detaches from the nozzle as the action of gravity overcomes the capillary retention forces. Hence the diameter of the drop is determined by the capillary length, and is well reproducible at $d$ = 2.6 $\pm$ 0.1 mm. The physical parameters of the liquid are: kinematic viscosity $\nu$ = 6.2 cSt, surface tension $\sigma$ = 0.066 N/m and density $\rho$= 1126 kg/m$^3$. These mixtures have a higher viscosity than water, which prevents both splashing - i.e. the splitting of the main drop into tiny ones - and corrugations at the border of the drop during the spreading phase. Hence during the spreading and the retraction, the drop mostly remains axi-symmetric. The height of fall $h$ prescribes the velocity at impact: $V_0 = (2 g h)^{1/2}$, which can be up to 5.5 m/s. Using backlighting together with a high-speed camera (at a maximal rate of 4700 frames/s, with a resolution of 576$\times$576), the shape of the interface during the spreading and bouncing processes can be determined. The magnification allows for an accuracy of about 15 $\mu$m per pixel. 

\begin{figure}
\begin{center}
\includegraphics[scale=0.095]{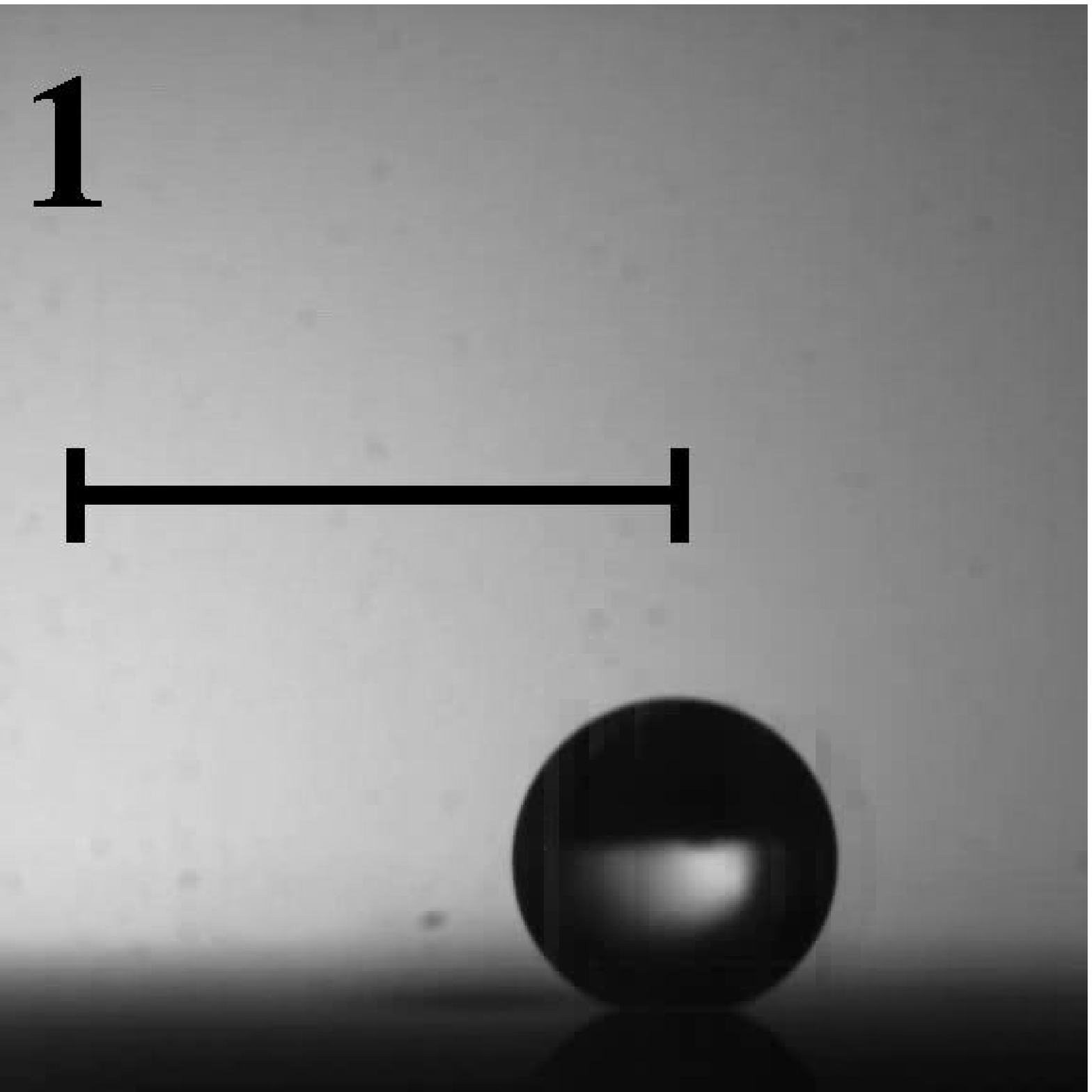}
\includegraphics[scale=0.095]{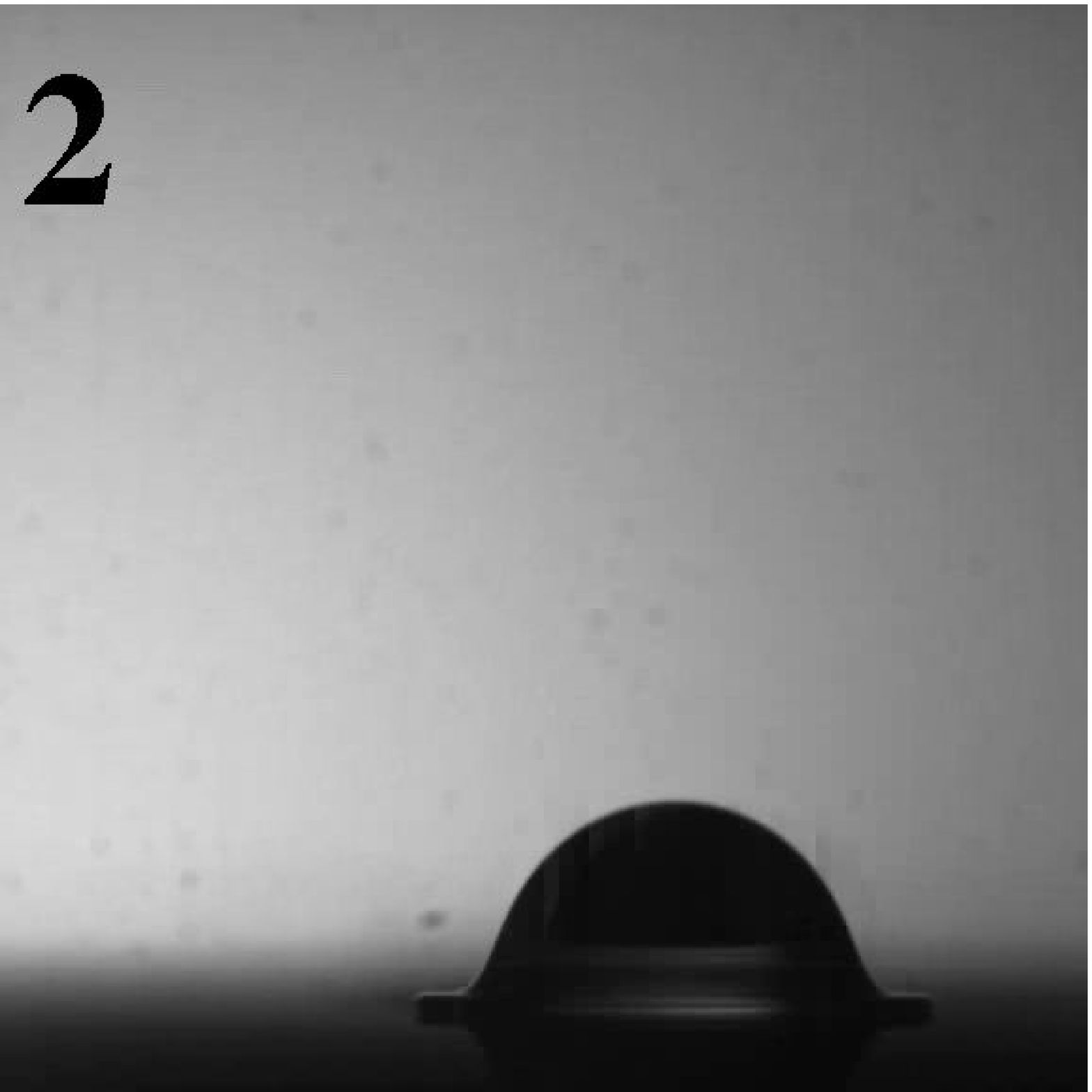}
\includegraphics[scale=0.095]{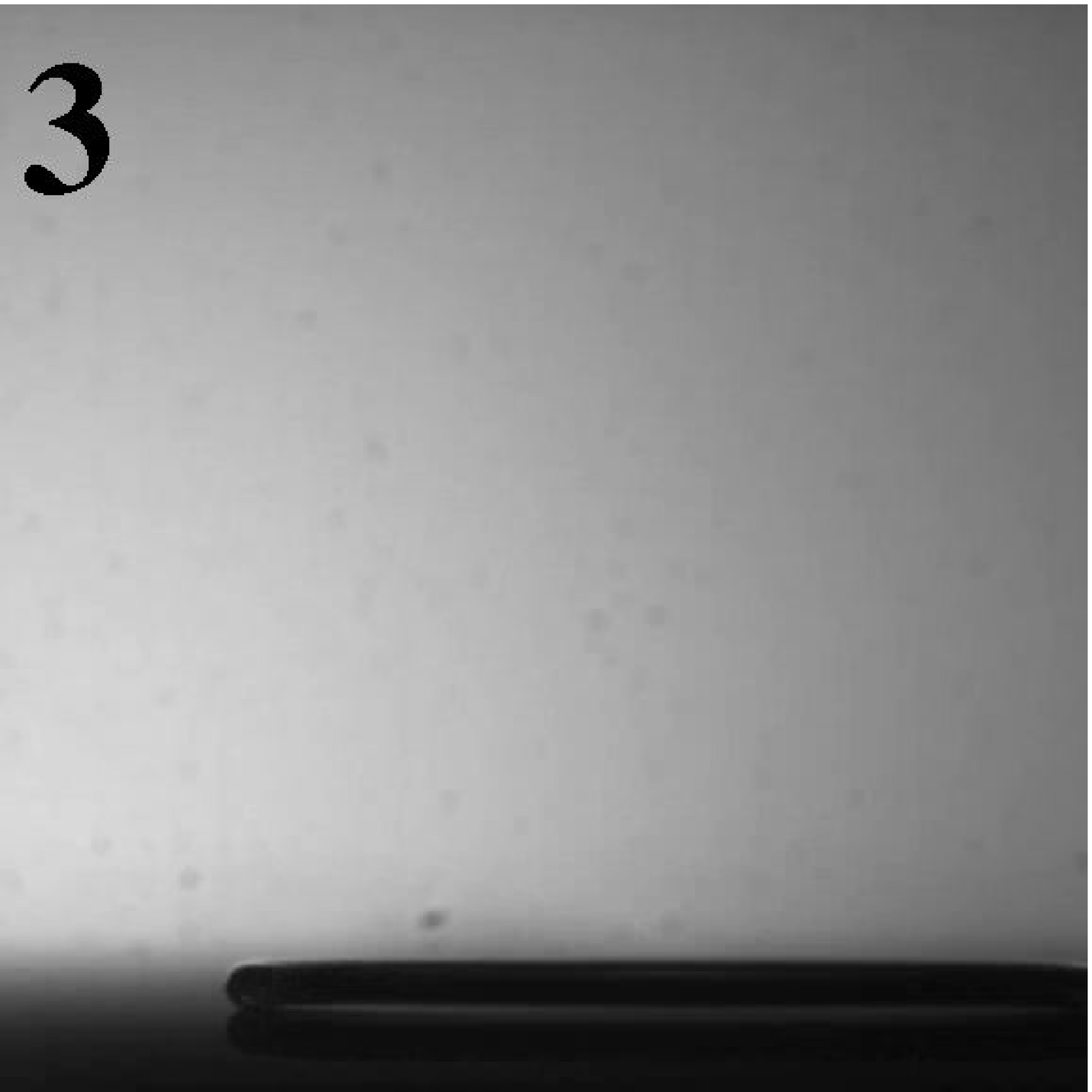}
\includegraphics[scale=0.095]{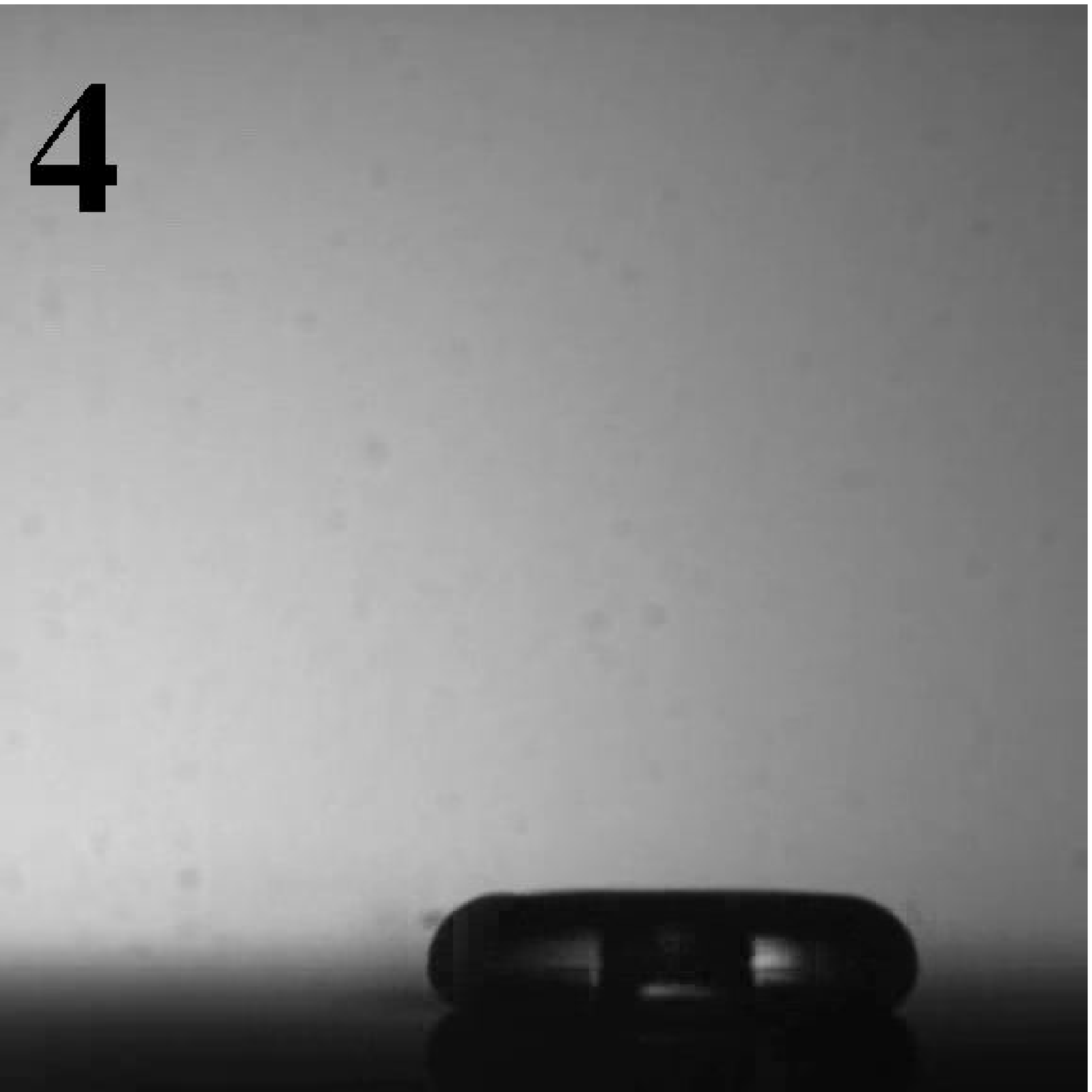}
\includegraphics[scale=0.095]{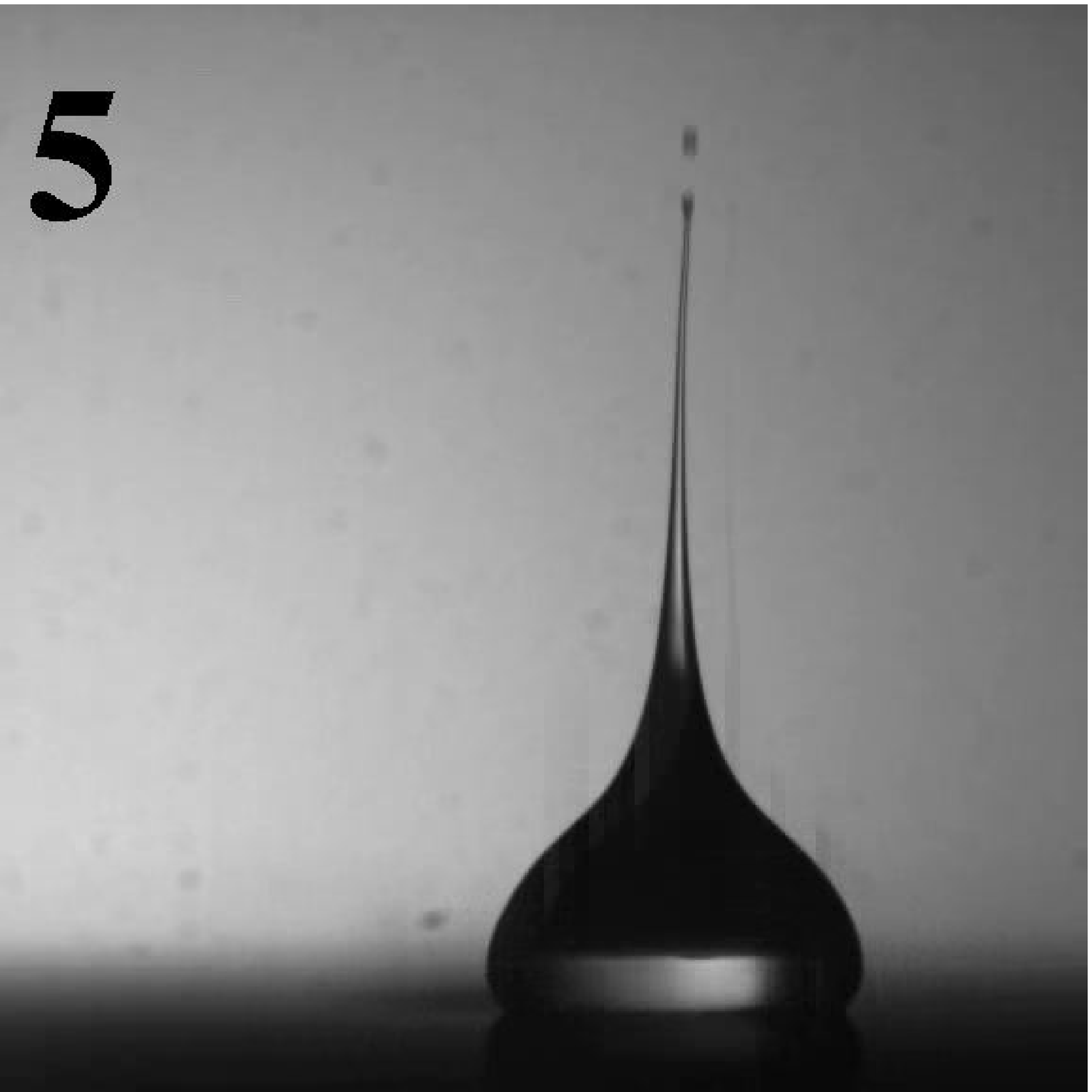}
\includegraphics[scale=0.095]{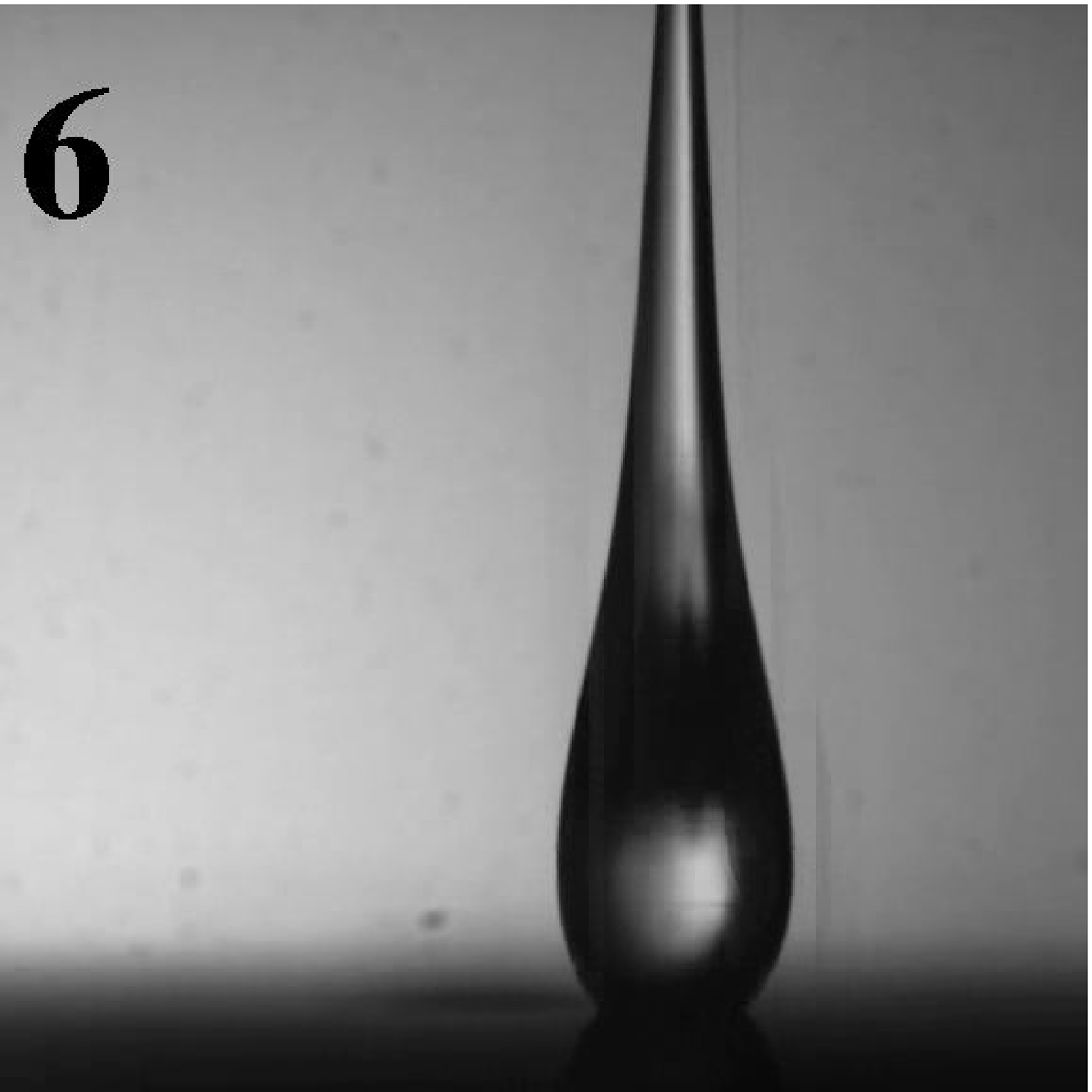}
\includegraphics[scale=0.095]{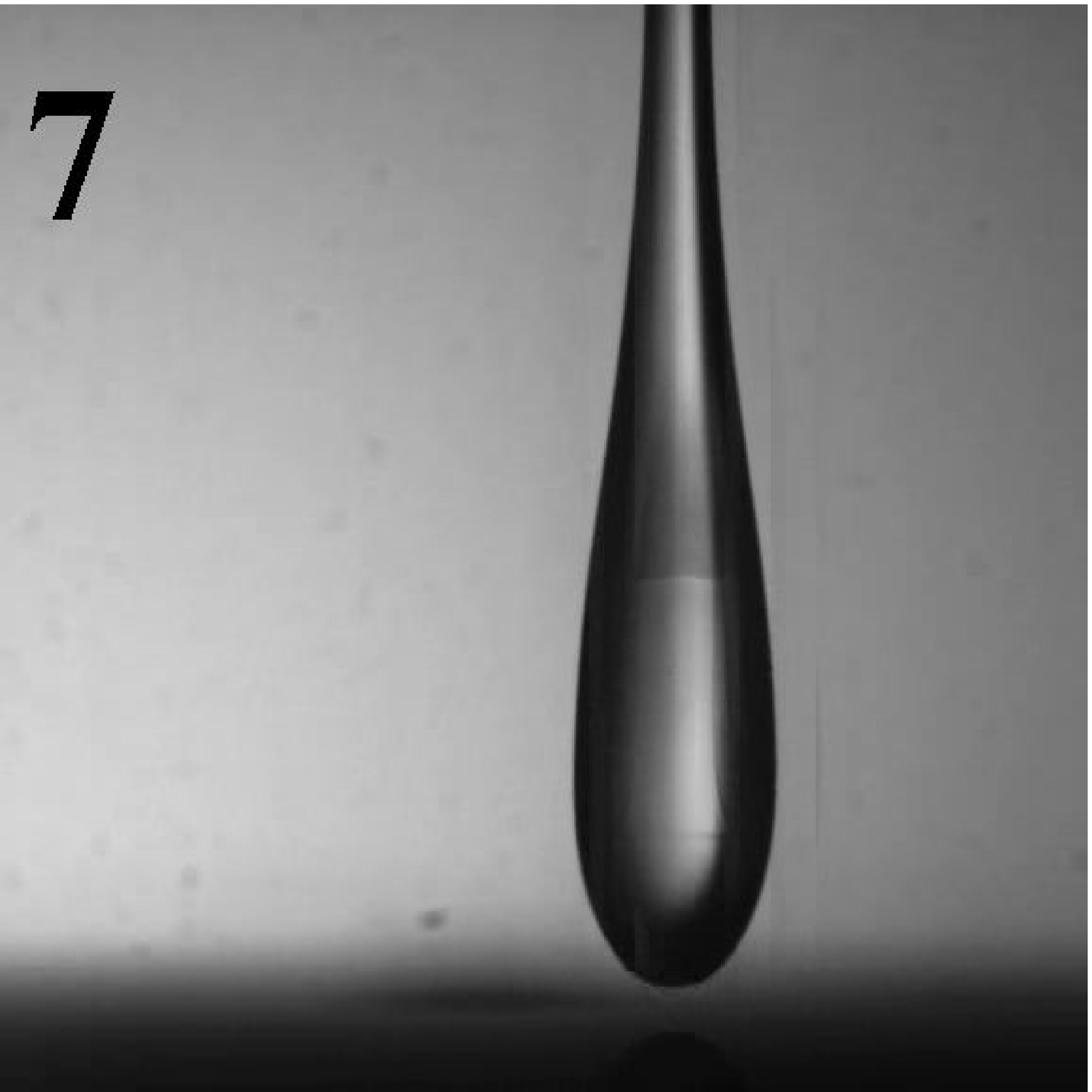}
\includegraphics[scale=0.095]{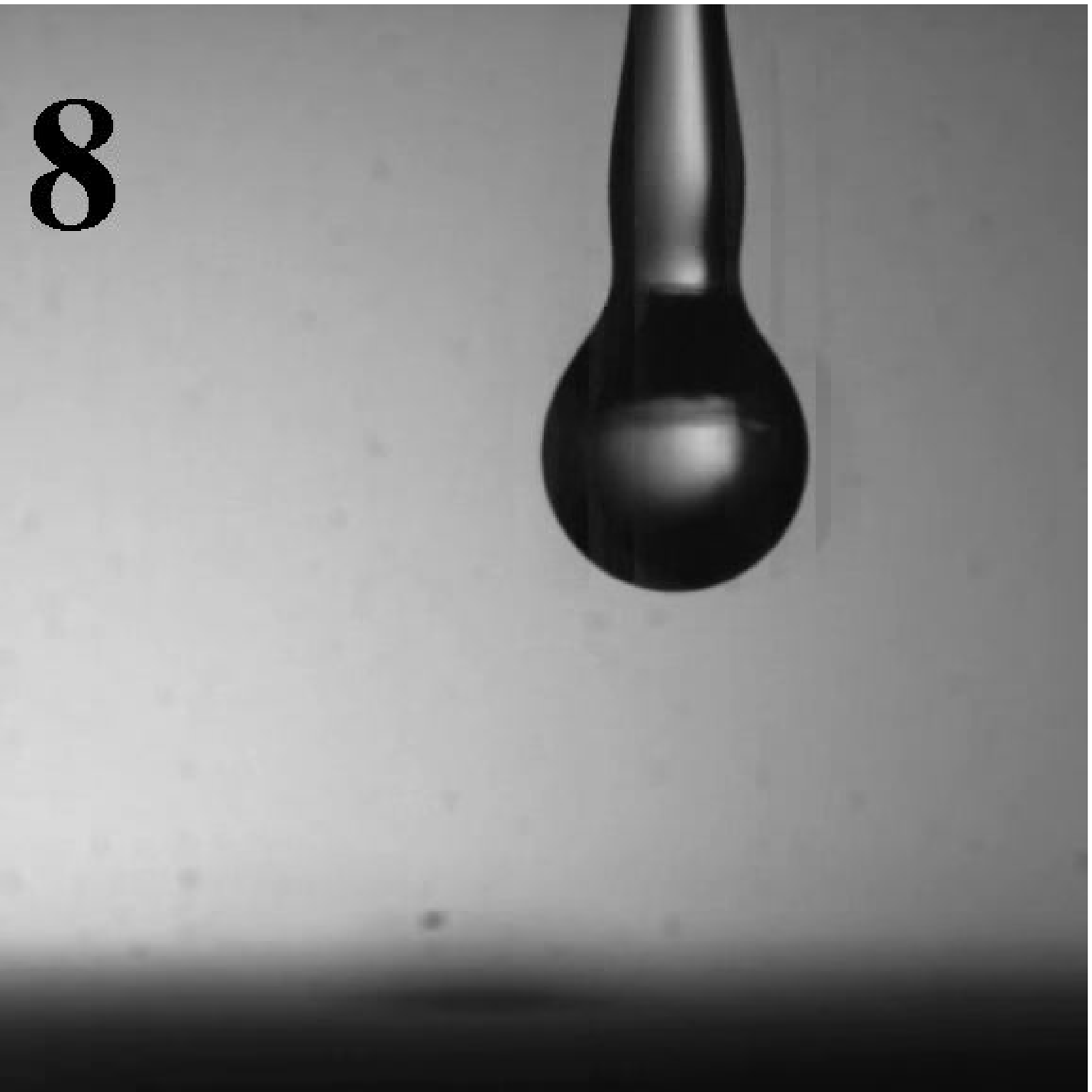}
\includegraphics[scale=0.4]{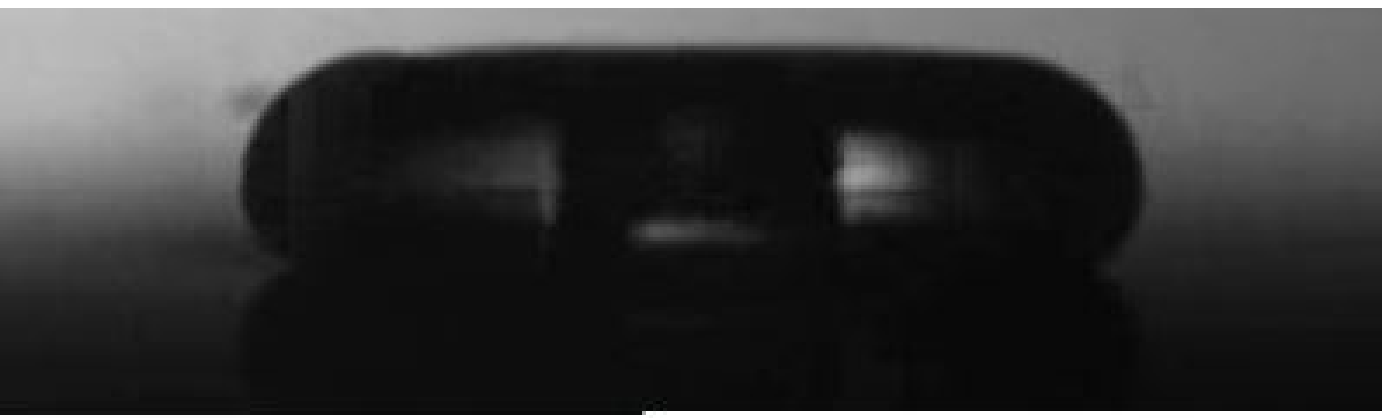}
\caption{\textit{Top} Successive shots showing the impact and the bouncing of a drop, without impalement (P3 substrate, $We$ = 600). The scale bar is 5 mm. The overall sequence lasts about 2.3 ms. \textit{Bottom} Close-up view of shot 4, showing an air cavity at the centre of the drop.}
\label{fig:no_impalement}
\end{center}
\end{figure}

\begin{figure}
\begin{center}
\includegraphics[scale=3]{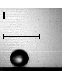}
\includegraphics[scale=3]{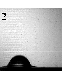}
\includegraphics[scale=3]{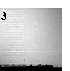}
\includegraphics[scale=3]{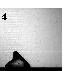}
\includegraphics[scale=3]{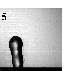}
\includegraphics[scale=3]{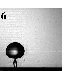}
\includegraphics[scale=3]{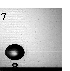}
\includegraphics[scale=3]{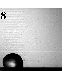}
\caption{Successive shots showing the impact and the bouncing of a drop, followed by an impalement (P1 substrate, $We$=727). The scale bar is 5 mm. The overall sequence lasts about 4.7 ms.}
\label{fig:impalement}
\end{center}
\end{figure}

In drop impact problems, the Weber (We) number, traditionally used to compare the influences of inertia and capillarity, is defined as:

\begin{equation}
We = \frac{\rho V^2 d}{\sigma}
\label{eq:weber}
\end{equation}

A typical rebound is shown in Fig.~\ref{fig:no_impalement}-\textit{Up}, where a drop impacts a P3 surface at $V_0$ = 3.68 m/s: the snapshots (1-8) show the successive phases of spreading where the drop's shape turns into a pancake (1-3), retraction when the drop when the radius of the pancake diminishes (4-5), bouncing when the drop takes off the surface (6-8). From direct observation of the movies, it is remarkable that no liquid has protruded the surface. This points out that the ultra-fine singular jet already observed by Bartolo \textit{et al.} \cite{Bartolo_prl} appears here for most of the velocities tested. The occurrence of this singular jet, due to the collapse of the internal air cavity visible in snapshot (4), is a signature of the absence of pinning (see Fig. \ref{fig:no_impalement}-\textit{Bottom}).

On the contrary, in Fig.~\ref{fig:impalement}, a drop impacts a P1 surface at $V_0$= 4.05 m/s. Although the spreading process (1-3) is similar to that of the previous situation (except the small corrugations at the periphery), the retraction phase (4-5) is dramatically modified. The bouncing (6-8) is very different from the previous case, as a droplet of liquid remains pinned on the surface (snapshot 7). Also it is striking to observe that the impalement is related to a dramatic decrease of the height of bouncing (see end of sequences on Fig. \ref{fig:no_impalement} and \ref{fig:impalement}). This suggests that there could be a massive energy dissipation process, due to the large value of the CAH, near the vicinity of the moving contact line during the spreading and the receding process. On the contrary, when the liquid does not penetrate the pores, the dissipation should only be due to the friction on top of the NWs, which constitutes a small effective area.

Thus, the best surfaces in terms of impalement threshold are those designed in the P3 process, corresponding to the double-layered network and the longest NWs. Under the accessible experimental conditions (speed at impact up to 5.5 m/s), it was not possible to provoke impalement on such surfaces. It means that such surfaces resist pressures larger than about 34 kPa. Here, the pressure is taken as equal to the dynamical pressure $P = \rho V^2$, similarly to the previous studies \cite{Bartolo06,Quere_Reyssat06}. For P3 substrates, the threshold pressure is much higher than previous surfaces consisting of micro-pillars: in \cite{Lafuma_Quere03}, the pressure threshold was about 200 Pa, in \cite{Bartolo06} the maximal threshold was around 380 Pa, and in \cite{Quere_Reyssat06} it was close to 10 kPa for the highest and narrowest pillars.

\begin{figure}
\begin{center}
(a)\includegraphics[scale=3.8]{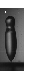}
\includegraphics[scale=3.8]{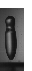}
(b)\includegraphics[scale=3.8]{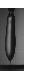}
\includegraphics[scale=3.8]{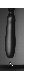}
(c)\includegraphics[scale=3.8]{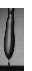}
\includegraphics[scale=3.8]{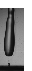}
\includegraphics[scale=3.8]{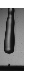}
\caption{Details of the impalement process, for a P2 surface. (a) $P \le P_c$. (b) $P \simeq P_c$. (c) $P \ge P_c$. The scale bar is 2 mm.}
\label{fig:detail_empal}
\end{center}
\end{figure}

With most of the surfaces tested except for the P3 surfaces (threshold was too high), the impalement threshold was determined accurately by the high-speed movies. Figure \ref{fig:detail_empal}-a,b,c shows three detailed sequences when the pressure at impact P is slightly below (a), equal to (b) and slightly larger (c) than $P_c$. The threshold coincides with the appearance of the tiniest observable drop (of diameter about 30 $\mu$m) just at the location of impact, where pressure is maximal. If the pressure at impact is strong enough to lead to total impalement (Wenzel state), then the drop is locally subjected to a large retention force. During bouncing, the retraction forces of capillary origin pull the liquid back. If the liquid is locally anchored in an irreversible way, the retraction is not strong enough to entrain the impaled liquid out of the texture. Hence, the liquid cylinder thins close to the surface and locally pinches off due to the Rayleigh-Plateau instability, leaving a droplet stuck on the surface.  For P1 and P2 surfaces, the threshold is equal respectively to 0.77 kPa and 15.9 kPa.

For surfaces P1, P2 and P3, the Weber number at impalement threshold is equal respectively to 30, 656 and 1340.

\textit{Electrowetting experiments} - In parallel, we carried out experiments of EW on each of the three different surfaces. At zero voltage, all the realized surfaces present a static 160 $\pm$ 3 $^{\circ}$ contact angle and a hysteresis of 0 $\pm$ 6 $^{\circ}$ leading to a 'rolling ball effect'.

The measurements are obtained in the following way: the voltage is applied between the drop and the substrate with a signal generator (CENTRAD GF 265, ELC, France), 2 to 20 Volts output at 1 kHz, coupled to a 50 dBm high-voltage amplifier (TEGAM, USA).
The voltage is varied between 10 to 135 Volts at the amplifier output. A goniometer (GBX, France) was used to record and measure the CA during EW. To determine the reproducibility of the EW phenomenon, the following method is applied: (1)Droplet deposition on the surface (3 $\pm$ 0.1 $\mu$l) through a metallic hydrophobic coated needle. (2) A unique voltage is applied (during 1 second) to the droplet through the same conductive needle. (3) Return at null voltage during 1 second. (4) Steps 2 and 3 are repeated ten times (EWOD cycle) at the same voltage.
For the P1 and P2 substrates, the EW shows irreversibility whatever the applied tension: the droplet CA dramatically decreased to a value depending of the voltage tension and stayed at this value when the voltage is turned off. Only the P3 substrate shows reversibility: the CA decreased as the voltage is applied and comes back to its original value obtained at zero voltage. The maximum contact angle variation under reversible conditions is 23$^{\circ}$ and this is obtained at 110 V, as already reported by Verplank et al. \cite{Verplank_etal07}. In order to understand the influence of EW on the impalement transition, we measured its receding and advancing contact angle after the step 4.

Once the voltage switched off after ten EW cycles, the droplet volume is either decreased (receding contact angle $\theta_r$) or increased (advancing contact angle $\theta_a$) respectively by sucking up or pumping up liquid. The operation is repeated at least twice at different position of the substrate for each voltage. Contact angle hysteresis is deduced from the difference between $\theta_a$ and $\theta_r$. P1 substrates show a high CAH (up to 70$^{\circ}$) after EW, which reveals the deep liquid impalement, as soon as the applied voltage is larger than about 15 V. For P2 substrates, the CAH increases dramatically above 50 V, and it ranges around 20 $\pm$ 3$^{\circ}$ beyond this threshold. P3 substrates display a low CAH - smaller than 5$^{\circ}$, for voltages up to 110 V. Above a threshold of 110 to 120 V, the CAH increases gently to 10$^{\circ}$, which shows that the drop begins to be impaled on the texture.

\begin{figure}
\begin{center}
\includegraphics[scale=0.6]{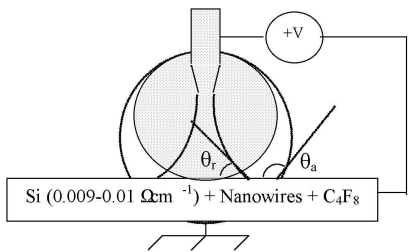}
\caption{Setup of contact-angle hysteresis measurement (see text for details).}
\label{fig:mes_hyst}
\end{center}
\end{figure}

\begin{figure}
\begin{center}
\includegraphics[scale=0.5]{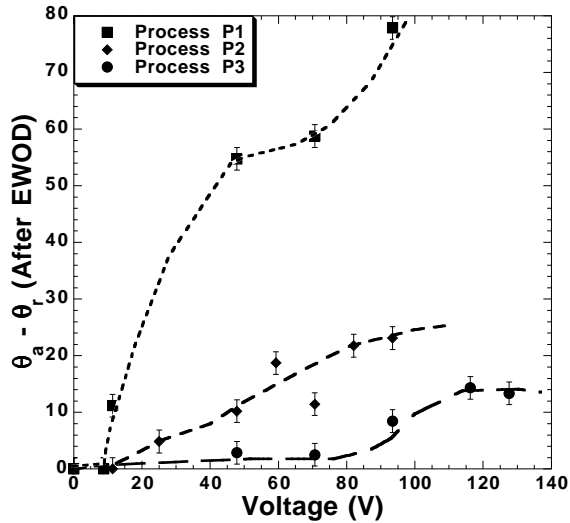}
\caption{Hysteresis $\theta_a - \theta_r$  as a function of applied tension of a droplet after 10 electrowetting cycles on three different SH surfaces. The plain, dashed and dotted lines are guides for the eye.}
\label{fig:hyst}
\end{center}
\end{figure}

The irreversible behaviour is appreciated upon two criterions: (1) the static CA after the cycles of EW does not go back to its initial value (160$ \pm 3 ^{\circ}$).  (2) the CAH after the cycles of EW overcomes a threshold value larger than the range of error in measurements, hence about 7 or 8 $^{\circ}$. These two criterions are independent, as it is possible to fill criterion (1) but not (2).

From these measurements, some conclusions are drawn: 

- the range of reversibility measured on various substrates during EW follows the same trend as the impalement threshold caused by a mechanical external pressure.

- the threshold for impalement/irreversibility increases with the length and density of NWs, while the value of CAH is smaller. This suggests that for multiple-layer substrates the impalement could only be partial, especially if the lower layers are denser than the upper ones. This would mean that partial impalement states do exist on such NWs carpets, as recently evidenced on micro-pillar arrays \cite{Moulinet_Bartolo}.

- the relevance of partial impalement is reinforced by the increase of CAH with the voltage. This would indeed imply that the impalement of liquid goes deeper and deeper, until it reaches the dense layer of crossed NWs.

- an additional observation is that the receding CA exhibits a jump during the receding contact-line phase (droplet suction), leading to a sudden decrease of CAH, for the three kind of substrates. Hence, the measurements of CAH were done just before these jumps. However, except for P3 substrates, the CAH after jump does not generally return to its value before EW. This suggests that the contact-line experienced depinning, and that the drop locally escaped from impalement - although still being partially impaled at its center.

The relationship between a high impalement threshold and a large range of reversibility in EW could be explained by the following reasons. First, as recently evidenced by Mugele and co-workers \cite{MugeleReview}, EW deforms the drop at scales larger than the thickness of the dielectric (here equal to 300 nm). Consequently, the equilibrium shape of the liquid-vapor interface inside the texture is still matching the Young's angle at zero voltage, and the capillary resistance to penetration should be the same for any voltage. Hence, the effect of EW  could be seen as additional electro-mechanical forces acting the same way as an external pressure.

\begin{figure}
\begin{center}
(a)\includegraphics[scale=0.15]{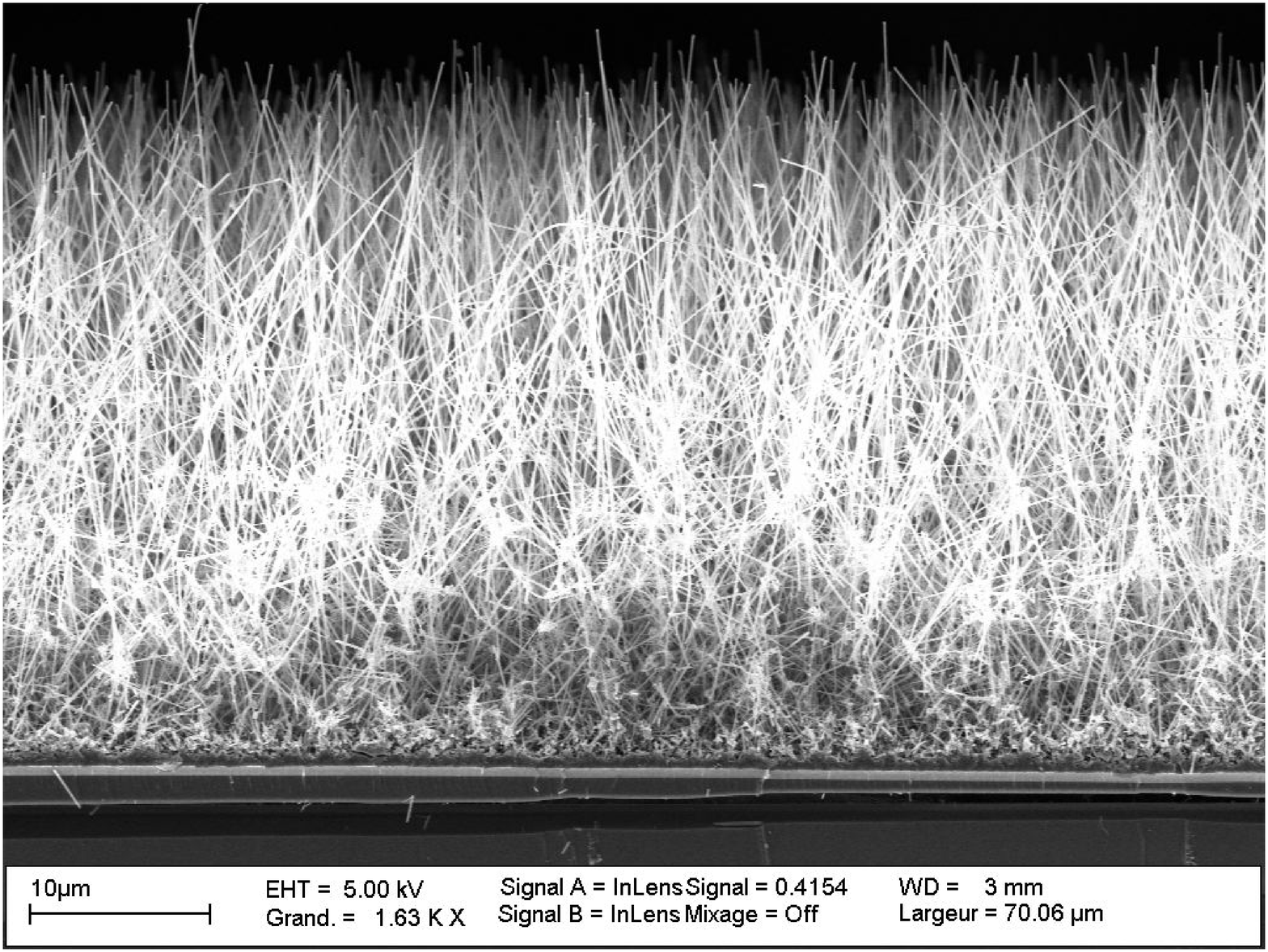}
(b)\includegraphics[scale=0.15]{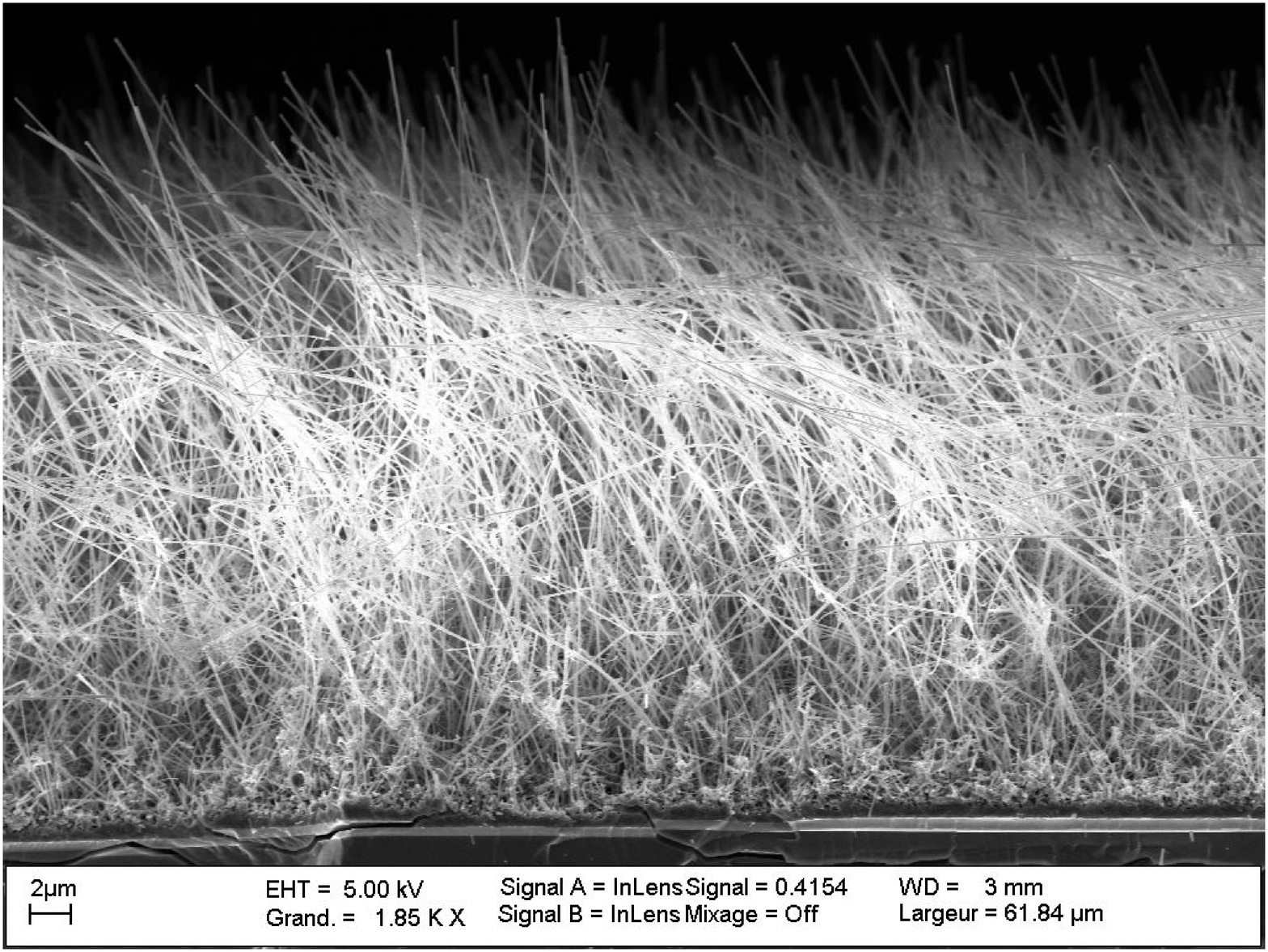}
\caption{SEM pictures of the P3 designed surfaces (a) before and (b) after impact.}
\label{fig:impact}
\end{center}
\end{figure}

This study also suggests future issues about the determination of the mechanical resistance of the NWs carpets. This issue is revealed by SEM visualizations at the location of impact, before and after impact. In Fig.~\ref{fig:impact}-(a) and (b), it is visible that NWs in the superficial layer bent over, whereas those in the deeper layers remained straight. Again, it looks like the impalement of liquid on such surfaces is only partial, and does not touch the bottom of the substrate. A possible interpretation could be the following: as far as the density of NWs ensures that the liquid does not percolate to the bottom of the surface, the impalement is only partial and hence could be reversible. This is consistent with the recent theoretical study by Kusumaatmaja et al. \cite{Kusumaatmaja08}: the lower layer of cross wires should prevent the curvature-induced mechanism for collapse.

\textit{Conclusions - Prospective studies} - This study performed on silicon nanowires surfaces shows the correlation between the resistance to impalement on both quasi-static EW and dynamical drop impact. The very high resistance of some of the surfaces (P3) seems to be related to both the length of the NWs and the specific structure of their network. Although this issue needs to be addressed more systematically, a first clue provided here is that the multi-layered structures can maintain the liquid in partial impalement - thus reversible - configurations. However, due to the disorder in the entanglement of the NWs, it is hard to address a relation with the surface energy in simple ways. 

Up to now, surfaces with a double texturation (micron-sized posts and NWs) were proposed as the most promising solutions for a high resistance to impalement \cite{Patankar04,Gao_McCarthy06,Bhushan_Nosonovsky06}. Following Patankar's initial point of view, in the very next future, we plan to test surfaces formed by micro-pillars covered with NWs, ensuring a double-scaled roughness. For the few surfaces tested, it is remarkable that the correlation between the EW and drop impact impalement thresholds is also observed. Also, it turns out that the gap between the two scales needs not to be too large. Then, a possible solution could be the use a third intermediate scale between the 10 $\mu m$-diameter micro-pillars and the 100 nm-diameter NWs.

\textit{Acknowledgments - } The authors thank S. Coudert for his kind support during the visualizations and P. Lefebvre-Legry for her technical support. The Centre National de la Recherche Scientifique (CNRS) is gratefully acknowledged for financial support.

\end{document}